\documentstyle[epsfig]{aprim10}
\input{epsf}

\newif\ifAMStwofonts

\ifoldfss
  \ifCUPmtlplainloaded \else
    \NewTextAlphabet{textbfit} {cmbxti10} {}
    \NewTextAlphabet{textbfss} {cmssbx10} {}
    \NewMathAlphabet{mathbfit} {cmbxti10} {} 
    \NewMathAlphabet{mathbfss} {cmssbx10} {} 
  \fi
  \ifAMStwofonts
    \ifCUPmtlplainloaded \else
      \NewSymbolFont{upmath} {eurm10}
      \NewSymbolFont{AMSa} {msam10}
      \NewMathSymbol{\upi}     {0}{upmath}{19}
      \NewMathSymbol{\umu}     {0}{upmath}{16}
      \NewMathSymbol{\upartial}{0}{upmath}{40}
      \NewMathSymbol{\leqslant}{3}{AMSa}{36}
      \NewMathSymbol{\geqslant}{3}{AMSa}{3E}

    \fi
  \fi
\fi 

\ifnfssone
  \newmathalphabet{\mathit}
  \addtoversion{normal}{\mathit}{cmr}{m}{it}
  \addtoversion{bold}{\mathit}{cmr}{bx}{it}
  \newmathalphabet{\mathbfit} 
  \addtoversion{normal}{\mathbfit}{cmr}{bx}{it}
  \addtoversion{bold}{\mathbfit}{cmr}{bx}{it}
  \newmathalphabet{\mathbfss} 
  \addtoversion{normal}{\mathbfss}{cmss}{bx}{n}
  \addtoversion{bold}{\mathbfss}{cmss}{bx}{n}
  \ifAMStwofonts
    \ifCUPmtlplainloaded \else
      \UseAMStwoboldmath
      \makeatletter
      \new@mathgroup\upmath@group
      \define@mathgroup\mv@normal\upmath@group{eur}{m}{n}
      \define@mathgroup\mv@bold\upmath@group{eur}{b}{n}
      \edef\UPM{\hexnumber\upmath@group}
      \new@mathgroup\amsa@group
      \define@mathgroup\mv@normal\amsa@group{msa}{m}{n}
      \define@mathgroup\mv@bold\amsa@group{msa}{m}{n}
      \edef\AMSa{\hexnumber\amsa@group}
      \makeatother
      \mathchardef\upi="0\UPM19
      \mathchardef\umu="0\UPM16
      \mathchardef\upartial="0\UPM40
      \mathchardef\leqslant="3\AMSa36
      \mathchardef\geqslant="3\AMSa3E
    \fi
  \fi
\fi 

\ifnfsstwo
  \DeclareMathAlphabet{\mathbfit}{OT1}{cmr}{bx}{it}
  \SetMathAlphabet\mathbfit{bold}{OT1}{cmr}{bx}{it}
  \DeclareMathAlphabet{\mathbfss}{OT1}{cmss}{bx}{n}
  \SetMathAlphabet\mathbfss{bold}{OT1}{cmss}{bx}{n}
  \ifAMStwofonts
    \ifCUPmtlplainloaded \else
      \DeclareSymbolFont{UPM}{U}{eur}{m}{n}
      \SetSymbolFont{UPM}{bold}{U}{eur}{b}{n}
      \DeclareSymbolFont{AMSa}{U}{msa}{m}{n}
      \DeclareMathSymbol{\upi}{0}{UPM}{"19}
      \DeclareMathSymbol{\umu}{0}{UPM}{"16}
      \DeclareMathSymbol{\upartial}{0}{UPM}{"40}
      \DeclareMathSymbol{\leqslant}{3}{AMSa}{"36}
      \DeclareMathSymbol{\geqslant}{3}{AMSa}{"3E}
    \fi
  \fi
\fi 

\ifCUPmtlplainloaded \else
  \ifAMStwofonts \else 
    \def\upi{\pi}
    \def\umu{\mu}
    \def\upartial{\partial}
  \fi
\fi

\title[Uncover Ultra-cool Dwarfs with Large Area Surveys]{Uncover Ultra-cool Dwarfs with Large Area Surveys}

\author[Z.H. Zhang et al.]
       {Z.H. Zhang$^{1,2}$, H. R. A. Jones$^{2}$, D. J. Pinfield$^{2}$, R. S. Pokorny$^{1}$ and Z. Han$^{1}$\\
        $^1$National Astronomical Observatories/Yunnan
    Observatory, Chinese Academy of Sciences, Kunming 650011,
    China\\
        $^2$Centre for Astrophysics Research, Science and Technology Research Institute, University of Hertfordshire,
        College Lane, Hatfield \\ AL10 9AB, U.K.}
\date{}

\pagerange{\pageref{firstpage}--\pageref{lastpage}}
\pubyear{2008}

\begin{document}

\maketitle

\label{firstpage}

\begin{abstract}
We selected brown dwarf candidates from the seventh Data Release of
the Sloan Digital Sky Survey (SDSS DR7) with new photometric
selection criteria based on a parameteriaztion of well-known L and T
dwarfs. Then we confirmed their status with SDSS spectra. The
candidates without SDSS spectra are cross matched in the Two Micron
All Sky Survey (2MASS) and the Fourth Data Release of the UKIRT
Infrared Deep Sky Survey (UKIDSS DR4). With the help of colors based
on SDSS, 2MASS and UKIDSS, we are able to estimate spectral types of
our candidates. We obtain reliable proper motions using positional
and epoch information downloaded direct from the survey databases.

\end{abstract}


\section{Introduction}
Brown dwarfs occupy the mass range between the lowest mass stars and
the highest mass planets. The central temperature of a brown dwarf
is not high enough to achieve stable hydrogen burning like a star,
but all brown dwarfs will undergo short periods of primordial
deuterium burning very early in their evolution. Since the first
discovery of an L dwarf  (GD165 B; Becklin \& Zuckerman 1988) and a
T dwarf (Gl229 B; Nakajima et al. 1995), the projects searching for
brown dwarfs have involved a number of large scale surveys, for
example, the Deep Near-Infrared Survey (DENIS; Epchtein et al.
1997), the Two Micron All Sky Survey (2MASS; Skrutskie et al. 2006)
and the Sloan Digital Sky Survey (SDSS; York et al. 2000;
Adelman-MaCarthy et al. 2008). Over five hundred L dwarfs and one
hundred T dwarfs have been found in large scale sky surveys in the
last decade (see, DwarfsArchives.org for a full list). Nearly 200 L
and T dwarfs have been found in SDSS (e.g. Fan et al. 2000; Geballe
et al. 2002; Hawley et al. 2002; Schneider et al. 2002; Knapp et al.
2004; Chiu et al. 2006), and more than 300 in 2MASS (e.g. Burgasser
et al. 1999, 2002, 2004; Kirkpatrick et al. 1999, 2000; Gizis et al.
2000; Cruz et al. 2003, 2007; Kendall et al. 2003, 2007; Looper et
al. 2007; Reid et al. 2008). More recently, the UKIRT Infrared Deep
Sky Survey (UKIDSS; Lawrence et al. 2007) is beginning to be very
effective in searching for T dwarfs (Kendall et al. 2007; Lodieu et
al. 2007; Warren et al. 2007; Burningham et al. 2008; Pinfield et
al. 2008) and has a strong potential to achieve the discovery of Y
dwarfs.

\section{Large Area Surveys}
The photometric data used in this work are from the SDSS DR7, the
Two Micron All Sky Survey and the Fourth Data Release of UKIDSS
Large Area Survey (LAS). The SDSS DR7 photometric data catalog
covers 11500 $deg^{2}$ in the main survey area (Legacy: 8417
$deg^{2}$, SEGUE: 3500 $deg^{2}$) in five bands (\emph{u}, \emph{g},
\emph{r}, \emph{i}, \emph{z}), with information on roughly 357
million distinct photometric objects (Legacy: 230 million, SEGUE:
127 million). Bright objects selected from the SDSS can be found in
the Two Micron All Sky Survey which have photometric data of objects
in \emph{J}, \emph{H}, and \emph{K} bands. Most of known ultra-cool
dwarfs found from SDSS have photometric data in 2MASS for they are
bright enough for 2MASS. But most our faint ultra-cool dwarf
candidates selected from SDSS could not be found in 2MASS. The
UKIDSS LAS will image an area of 4000 $deg^{2}$ at high Galactic
latitudes in the \emph{Y, J, H} and \emph{K} filters to $Y=20.5,
J=20.0, H=18.8, K=18.4$ which are about three magnitudes deeper than
2MASS in \emph{J, H} and \emph{K} band. The LAS target 4000
$deg^{2}$ is a subsection of the Sloan survey. The fourth Data
Release of UKIDSS LAS covers 1200 $deg^{2}$.

\section{Colors of Ultra-cool dwarfs}
Brown dwarfs are infrared objects and very faint in optical bands.
The SDSS $i-z$ color is particularly useful for L dwarf selection
(as first pioneered by Fan et al. 2000), and expanded on by others
(e.g. via the $i$-band drop-out method; e.g. Chiu et al. 2006). For
the cooler $T_{\rm eff}$ T dwarfs, almost all of the radiation is
emitted beyond 10000 {\AA}, and as such these objects are optically
much fainter than L dwarfs. SDSS is thus significantly less
sensitive to T dwarfs than to L dwarfs, but has the sensitivity to
identify L dwarfs out to distances beyond 100pc.

\begin{figure} 
   \centering
 \includegraphics[width=9.5cm]{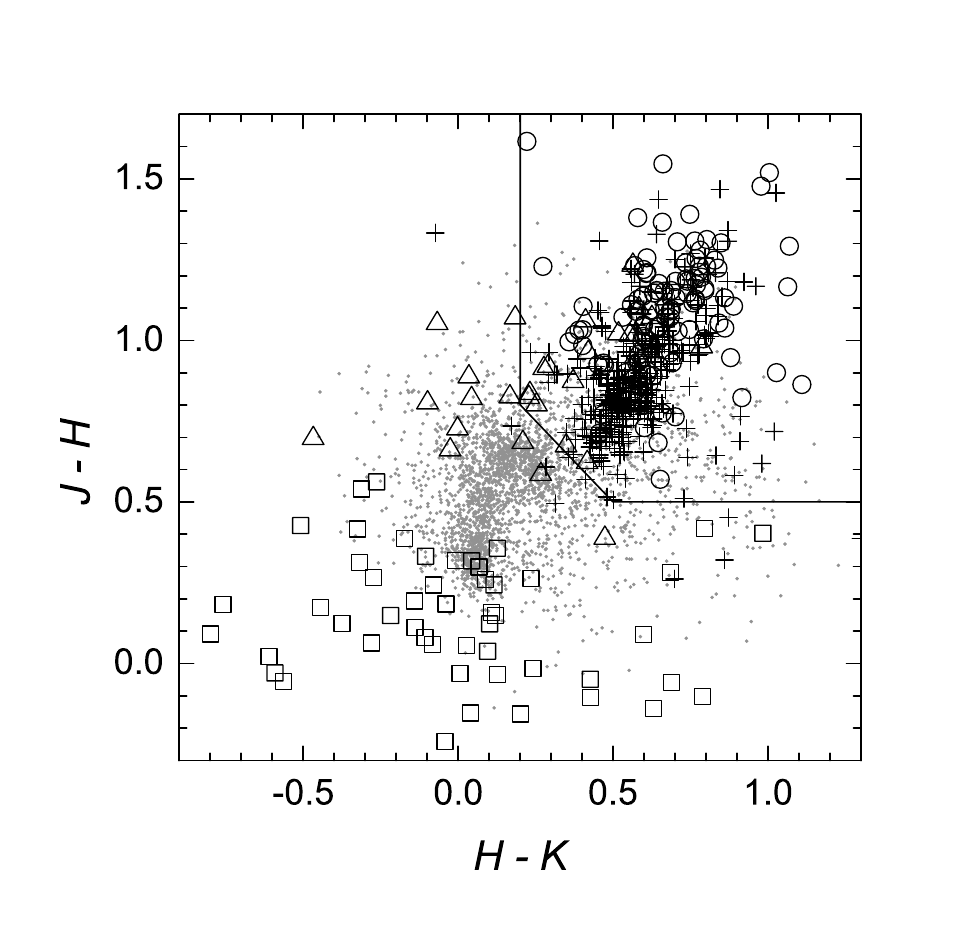}
 \caption[]{$J-H$ vs. $H-K$ diagram for previously identified L and T dwarfs.
Previously identified L0-L4.5 (\emph{crosses}), L5-L9.5 (\emph{open
circles}), T0-T3.5 (\emph{open triangles}) and T4-T7 (\emph{open
squares}) dwarfs. Note that the 2MASS criteria (solid lines) are
only applied when selecting photometric candidates, and not when
SDSS spectra are available. For comparison, the plot shows 2800
sources (\emph{dots}) taken from 3.14 $deg^2$ of 2MASS sky.}
\end{figure}
\begin{figure} 
\includegraphics[width=9.5cm]{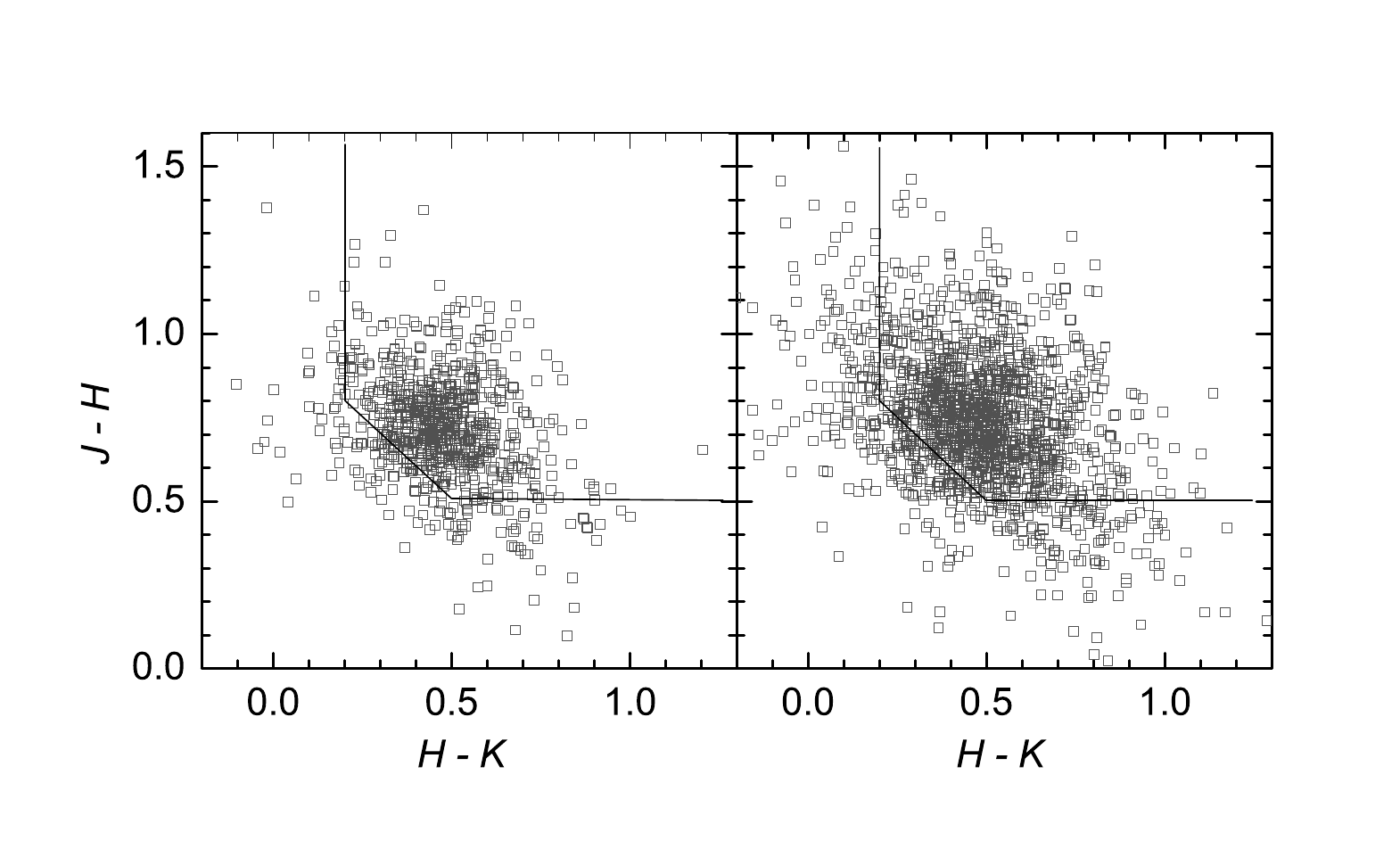}
 \caption[]{$J-H$ vs. $H-K$ diagram for
objects matched in 2MASS. Objects in the right hand panel have SDSS
red optical spectra. Objects in the left hand panel have no SDSS
spectra. Solid lines show our near infrared photometric selection
criteria in \emph{JHK} color-space }
\end{figure}

\begin{figure} 
\includegraphics[width=9cm]{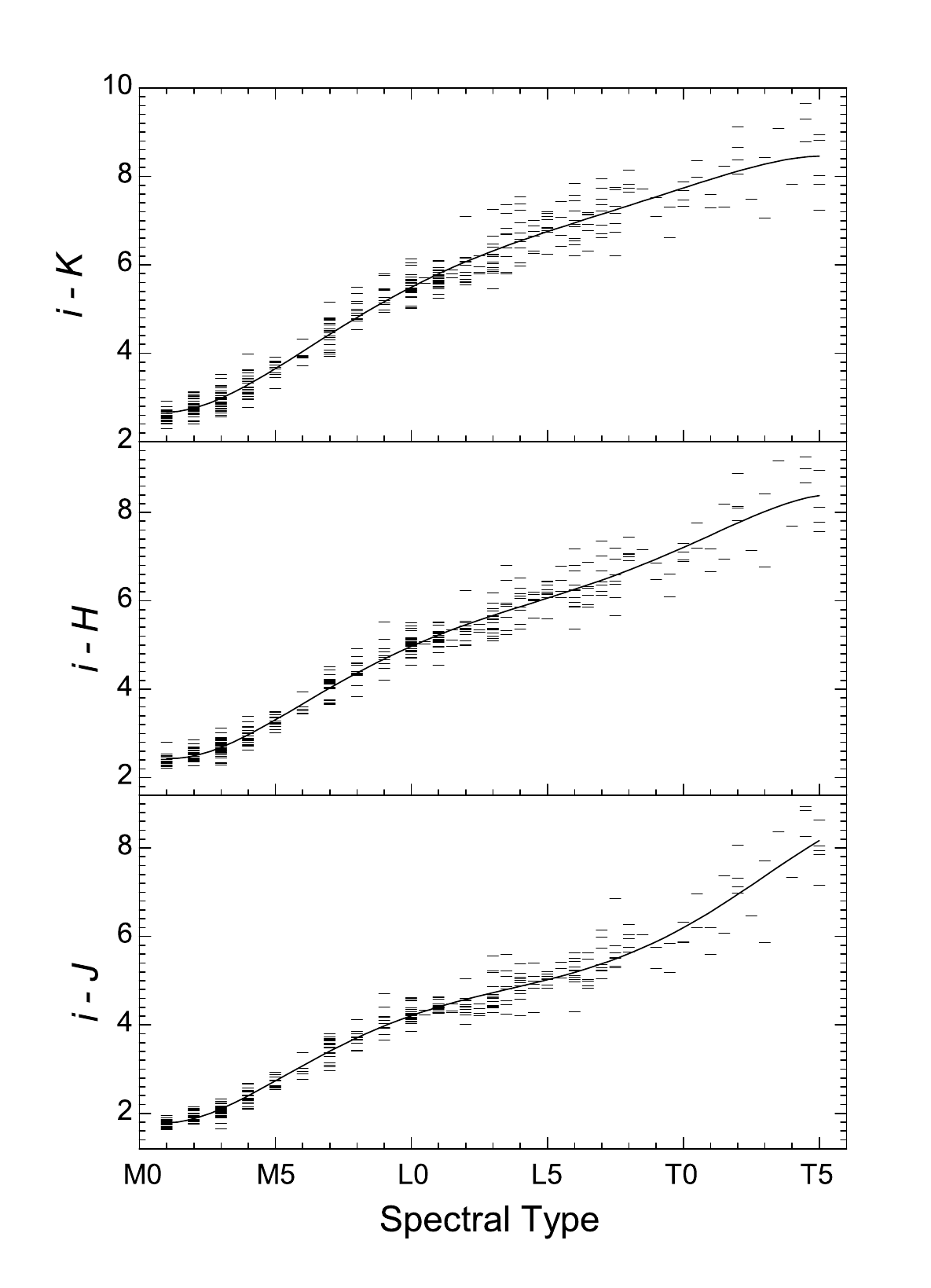}
 \caption[]{Polynomial fitting for
    color-spectral type relationships are indicated with a solid line.}
\end{figure}

\begin{figure} 
\includegraphics[width=9.5cm]{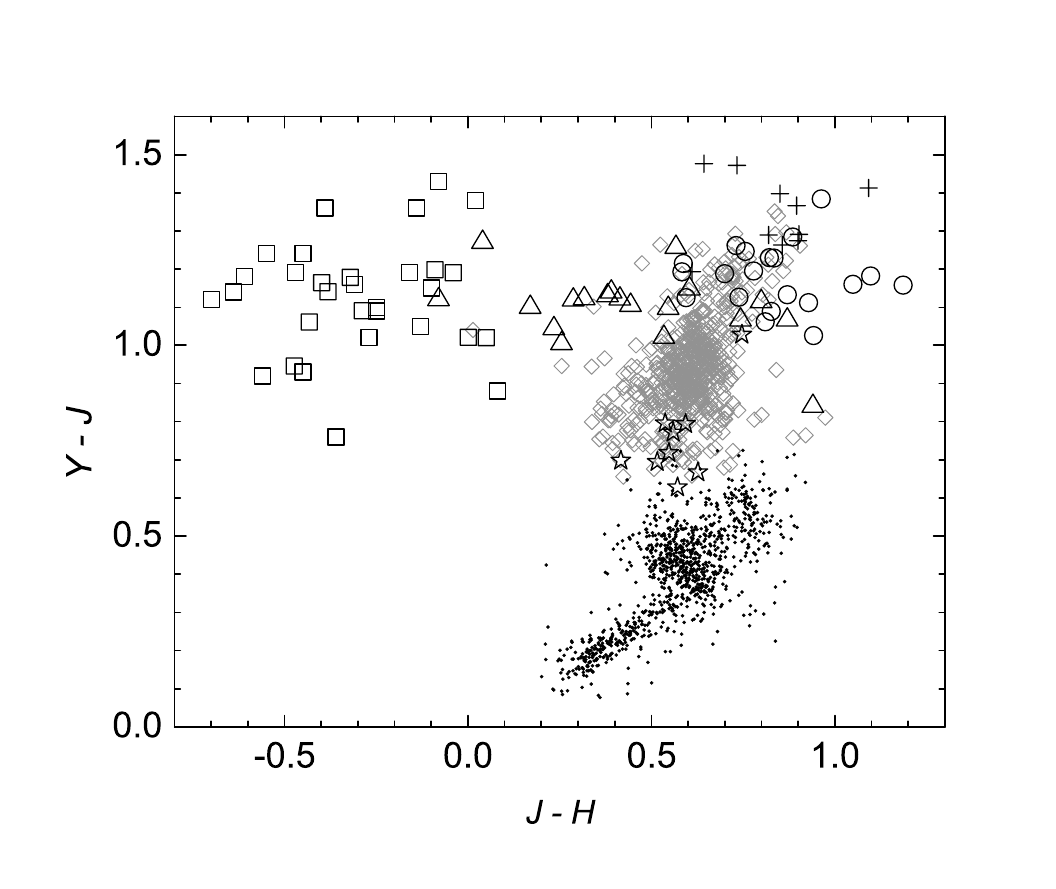}
 \caption[]{$Y-J$ vs. $J-H$ diagram for
known M, L and T dwarfs and new ultracool dwarf candidates
(\emph{open diamonds}) which matched in UKIDSS. Symbols are as in
Figure 1. For comparison, the plot shows 1024 sources (\emph{dots})
taken from UKIDSS LAS in 1 $deg^2$ with $Y < 18.5$.}
\end{figure}

We have made a study of L and T dwarf color-color parametre-space
using previously identified L and T dwarfs with photometric data
available from either SDSS or 2MASS (from DwarfsArchives.org, as of
September 25, 2007). A total of 431 L and 84 T dwarfs have 2MASS
photometric data (\emph{J}, \emph{H}, \emph{K}), and 193 L and 46 T
dwarfs have SDSS photometric data (\emph{u}, \emph{g}, \emph{r},
\emph{i}, \emph{z}). The study shows L dwarfs with a color of
$i-z>1.7$ and T dwarfs with a color of $i-z>3$. (Zhang et al. 2009).
Figure 1 shows 2MASS \emph{J-H} and \emph{H-K} color of L and T
dwarfs, solid lines shows the boundary of L dwarfs in \emph{JHK}
color space. We can see that the \emph{J-H }and \emph{H-K} colors
are red for L dwarfs and blue for late T dwarfs.

Around 1800 SDSS color selected late M and early L dwarfs are
matched in 2MASS catalogue (Zhang et al. in prepare). Figure 2 shows
the position of our SDSS color selected L and T dwarf candidates in
2MASS \emph{JHK} calor space, solid lines shows the boundary of L
dwarfs the same as in Figure 1.

We can estimate the spectral types of our ultra-cool dwarf
candidates with the SDSS and 2MASS data based on the relationships
between spectral types and SDSS-2MASS cross colors (e.g. Hawley et
al. 2002). With a large number of M, L and T dwarfs now available,
we made a study of the relationships between spectral types of M, L
and T dwarfs and their colors from SDSS and 2MASS. Figure 3 shows
the polynomial fitting for color-spectral type relationships.
SDSS-2MASS cross colors \emph{i-J, i-H} and \emph{i-K} are the best
colors for spectral typing.

Almost all the SDSS selected candidates can be detected in UKIDSS
LAS. Around 1000 out of 6000 late M and L dwarf candidates selected
from SDSS matched in UKIDSS. Figure 4 shows known M, L and T dwarfs
and 547 candidates matched in the Fourth Data Release of UKIDSS LAS.
As a M and L dwarf candidates population, our SDSS selected
candidates are match very well in 2MASS \emph{JHK} color space and
UKIDSS \emph{YJH} color space.


\section{Discussion}
SDSS-UKIDSS color-spectral type relationships yield much better
estimates for spectral of ultra-cool dwarfs relative to SDSS-2MASS
colors. With the increase in coverage of UKIDSS and the number of
ultra-cool dwarfs discovered, we can get enough ultra-cool dwarfs
with UKIDSS data to build new color-spectral type relationships.

The UKIDSS LAS target area is a subsection of SDSS. SDSS and UKIDSS
have high resolution images (0.4$^{\prime\prime}$pixel$^{-1}$ for
SDSS; 0.4$^{\prime\prime}$pixel$^{-1}$ for \emph{Y, H, K} and
0.2$^{\prime\prime}$pixel$^{-1}$ for UKIDSS \emph{J}) relative to
2MASS (1$^{\prime\prime}$pixel$^{-1}$) and have a epoch difference
of $\sim$3-8 years. So proper motions of ultra-cool dwarfs based on
SDSS and UKIDSS offer great potential to confirm faint L dwarfs as
UKIDSS coverage increases.

\section*{Acknowledgment}
Funding for the SDSS and SDSS-II has been provided by the Alfred P.
Sloan Foundation, the Participating Institutions, the National
Science Foundation, the U.S. Department of Energy, the National
Aeronautics and Space Administration, the Japanese Monbukagakusho,
the Max Planck Society, and the Higher Education Funding Council for
England. The SDSS Web Site is http://www.sdss.org/.

The SDSS is managed by the Astrophysical Research Consortium for the
Participating Institutions. The Participating Institutions are the
American Museum of Natural History, Astrophysical Institute Potsdam,
University of Basel, University of Cambridge, Case Western Reserve
University, University of Chicago, Drexel University, Fermilab, the
Institute for Advanced Study, the Japan Participation Group, Johns
Hopkins University, the Joint Institute for Nuclear Astrophysics,
the Kavli Institute for Particle Astrophysics and Cosmology, the
Korean Scientist Group, the Chinese Academy of Sciences (LAMOST),
Los Alamos National Laboratory, the Max-Planck-Institute for
Astronomy (MPIA), the Max-Planck-Institute for Astrophysics (MPA),
New Mexico State University, Ohio State University, University of
Pittsburgh, University of Portsmouth, Princeton University, the
United States Naval Observatory, and the University of Washington.

This work is based in part on data obtained as part of the UKIRT
Infrared Deep Sky Survey. This publication makes use of data
products from the Two Micron All Sky Survey. This research has made
use of the VizieR catalogue access tool, CDS, Strasbourg, France.
Research has benefitted from the M, L, and T dwarf compendium housed
at DwarfArchives.org and maintained by Chris Gelino, Davy
Kirkpatrick, and Adam Burgasser. This work was part supported by the
Natural Science Foundation of China under Grant Nos 10521001,
10433030 and the CAS Research Fellowship for International Young
Researchers.

\label{lastpage}

\clearpage

\end{document}